\documentclass[a4paper,11pt]{article}
\pdfoutput=1
\usepackage{jcappub}

\usepackage{color}
\usepackage{amsfonts}
\usepackage{graphicx}
\usepackage{amssymb}
\usepackage{amsmath}

\usepackage{eufrak}
\usepackage{mathrsfs}

\renewcommand{\thesection}{\arabic{section}}

\def\laq{~\raise 0.4ex\hbox{$<$}\kern -0.8em\lower 0.62ex\hbox{$\sim$}~}
\def\gaq{~\raise 0.4ex\hbox{$>$}\kern -0.7em\lower 0.62ex\hbox{$\sim$}~}

\def\beq{\begin{equation}}
\def\eeq{\end{equation}}
\def\bea{\begin{eqnarray}}
\def\eea{\end{eqnarray}}

\def \pa {\partial}

\def \ra {\rightarrow}
\def \la {\lambda}

\def \b {\beta}
\def \a {\alpha}

\def \Ga {\Gamma}
\def \ga {\gamma}

\def \da {\delta}

\def \r {\rho}

\def \up {\Upsilon}
\def \wt {\widetilde}

\title{Observation angles, Fermi coordinates, and the Geodesic-Light-Cone gauge}

\author[a]{G. Fanizza,}
\author[b]{M. Gasperini,}
\author[c]{G. Marozzi,}
\author[d]{G. Veneziano}

\affiliation[a]{Istituto Nazionale di Fisica Nucleare, Sezione di Pisa,\\
Largo B. Pontecorvo 3, 56127 Pisa, Italy\\
Center for Theoretical Astrophysics and Cosmology, Institute for Computational Science, \\
University of Zurich, Winterthurerstrasse 190, CH-8057, Zurich, Switzerland}
\affiliation[b]{
Dipartimento di Fisica, Universit\`a di Bari, 
Via G. Amendola 173, 70126 Bari, Italy,\\
Istituto Nazionale di Fisica Nucleare, Sezione di Bari, Italy,\\
Via G. Amendola 173, 70126 Bari, Italy
}
\affiliation[c]{
Dipartimento di Fisica, Universit\`a di Pisa, Largo B. Pontecorvo 3, 56127 Pisa, Italy,\\
Istituto Nazionale di Fisica Nucleare, Sezione di Pisa, Italy,\\
Largo B. Pontecorvo 3, 56127 Pisa, Italy
}
\affiliation[d]{CERN, Theory  Department, CH-1211 Geneva 23, Switzerland,\\
Coll\`ege de France, 11 Place M. Berthelot, 75005 Paris, France}

\emailAdd{giuseppe.fanizza@pi.infn.it}
\emailAdd{gasperini@ba.infn.it}
\emailAdd{Giovanni.Marozzi@unipi.it}
\emailAdd{Gabriele.Veneziano@cern.ch}

\abstract{We show that the angular directions locally measured by a static geodesic observer in a generic cosmological background and expressed in the system of Fermi Normal Coordinates always coincide with those expressed in the Geodesic-Light-Cone (GLC) gauge, up to a local transformation which exploits the residual gauge freedom of the GLC coordinates. This is not the case for other gauges -- like, for instance, the synchronous and longitudinal gauge -- commonly used in the context of observational cosmology.
We also make an explicit proposal for the GLC  gauge-fixing condition that ensures a full identification of its angles with the observational ones.}

\keywords{Fermi normal coordinates, geodesic-light-cone gauge, theoretical cosmology
 
\vskip13pt plus8pt minus11pt

\noindent{\bfseries\large\sffamily{Preprints:}} BA-TH/716-18, CERN-TH-2018-253
}

\begin{document}

\maketitle

%%%%%%%%%%%%%%%%%%%%%%%%%%
%%%%%%%%%%%%%%%%%%%%%%%%%%
\section{Introduction}
\label{Sec1}
\setcounter{equation}{0}
%%%%%%%%%%%%%%%%%%%%%%%%%%

One of the most important problems, when comparing theory with observational results, is to correctly identify the observed quantities with the mathematical variables used to formulate the theoretical predictions. In this paper we discuss this problem for the case of the angular variables appearing in all models concerning -- for instance -- a precise description of light propagation in perturbed cosmological backgrounds.

We shall assume, as usual, that the angles directly related to local observations are those measured by a free-falling observer, and we will identify such angles with those of the so-called system of Fermi Normal Coordinates (FNC) \cite{1,2}. In that system, indeed, the spacetime metric is locally flat around all points of a given worldline, with  corrections starting at quadratic order in  the distance from the wordline and  weighted by the components of the Riemann tensor (see e.g. \cite{3} for a recent discussion). 

Choosing in particular the worldline of a static geodesic observer, at rest in the so-called comoving frame, as the typical representation of a reference cosmic observer\footnote{This is not an absolute necessity but can be seen as the most useful convention to be used in order to compare calculations made in different coordinate systems. These should agree if one computes the  {\it same} physical observable as measured by the {\it same} physical observer. One of us (GV) would like to thank Misao Sasaki for a discussion on this issue.}, we will show that the angular directions of the corresponding FNC system can always be made to coincide with those of the so-called Geodesic-Light-Cone (GLC) gauge \cite{4}, modulo a redefinition that exploits the residual gauge freedom of the GLC coordinates. Such an identification of the angular coordinates with those of the FNC system turns out to be impossible in general in other gauges, like (for instance) the synchronous gauge.

The above results will be discussed for a general class of (inhomogeneous, anisotropic) cosmological backgrounds, and then illustrated for the particular example of a Bianchi-I type geometry. Such results are new to the best of our knowledge, and provide, in our opinion, additional (important) motivations supporting  the use of GLC coordinates for the computation of the relevant cosmological observables. We may recall, in particular, that both the red-shift  \cite{4} and the Jacobi map  \cite{5} (hence, luminosity and angular distance) take a particularly simple form in these coordinates.

In Sec. \ref{Sec2} we show that FNC and synchronous-gauge angles cannot be made to coincide for a generic metric, and illustrate this result in the case of a Bianchi I (anisotropic) geometry. In Sec. \ref{Sec3} we compare angles in the FNC and GLC gauge to lowest order in the distance from the observer's geodetic and show that,  unlike in the previous case, the identification is always possible by exploiting the residual gauge freedom of the GLC coordinates. The connection is explicitly worked out  in the case of a Bianchi-I metric. In Sec. \ref{Sec4} we summarize our main results, and propose a necessary and sufficient criterion for defining, {\it within the GLC metric}, a full observational gauge in which both time and angles agree with those of the FNC system.  Finally, in the Appendix we show how the considerations of Sec. \ref{Sec3} can be extended to next order in the distance from the observer's geodesic, while remaining all the time in the GLC gauge.

%%%%%%%%%%%%%%%%%%%%%%%%%%%%
%%%%%%%%%%%%%%%%%%%%%%%%%%%
\section{Angular directions in the Fermi and synchronous gauge}
\label{Sec2}
\setcounter{equation}{0}
%%%%%%%%%%%%%%%%%%%%%%%%%%%
%%%%%%%%%%%%%%%%%%%%%%%%%%%%

Let us consider the class of cosmological geometries which can be described, in the synchronous gauge (SG) and in Cartesian coordinates $x^\mu= (t, x^i)$, by six arbitrary degrees of freedom parametrized by the metric
\beq
g_{00} = -1, ~~~~~~~~~~~~~~ g_{0i}=0, ~~~~~~~~~~~~~~~~ g_{ij}=g_{ij} (t, \vec x).
\label{21}
\eeq
Let us consider a static geodesic observer located at the origin $x^i=0$, with four-velocity $u^\alpha=(1, \vec 0)$, and let us call $x^{\prime \alpha}= (t', x^{\prime A})$ the FNC system centered around the worldline of such a free-falling observer (for convenience, we will denote with capital Latin indices,  $A, B = 1,2,3$, the spatial components of the Fermi coordinates).

In the FNC system the metric is flat, up to corrections which are of second order in the spatial distance from the central geodesic \cite{1,2,3}. Choosing the $x^{\prime \alpha}$ coordinates in such a way that the geodesic is located at $x^{\prime  A}=0$, we can expand the FNC metric as
\beq
g'_{\alpha\beta}(x')= \eta_{\alpha\beta} +f_{\alpha A\beta B} (t')x^{\prime  A}x^{\prime B}+ \cdots ,
\label{22}
\eeq
where $\eta_{\a\b}$ is the Minkowski metric. The coefficients  $f_{\alpha A\beta  B} $  depend on the second derivatives of the metric, and are  related to the components of the Riemann tensor in Fermi coordinates, $R'_{\mu\nu\a\b}(x')$, evaluated at $x^{\prime  A}=0$ \cite{1,2,3}, by\footnote{We are using the metric and curvature conventions of \cite{2}.}:
\beq
f_{0A0B} = \left(R'_{0A0B}\right)_{x'^A=0}~;~~~~f_{0ACB} = \frac23 \left(R'_{0ACB}\right)_{x'^A=0}~;~~~~f_{CADB} = \frac13\left(R'_{CADB}\right)_{x'^A=0} .
\label{Riem}
\eeq
 The corresponding coordinate transformation $x^\mu \ra x^{\prime \alpha}$, connecting the synchronous and FNC system, can then be expanded around the geodesic as follows:
\beq
t'= \a(t)+ \b_k (t)x^k + \ga_{kl}(t) x^k x^l + \cdots, ~~~~~~~~~
x^{\prime A}= \a^A\,_k(t) x^k+ \b^A\,_{kl}(t) x^k x^l+\cdots .
\label{23}
\eeq
In the same limit ($x^i, x'^A \ra 0$) it is also convenient to expand the spatial part of the general synchronous metric (\ref{21}), up to second order, as
\beq
g_{ij}(t,x)=A_{ij}(t)+ B_{ijk}(t)x^k+ C_{ijkl}(t)x^k x^l + \cdots,
\label{24}
\eeq

The time-dependent coefficients appearing in the transformation equation (\ref{23}) can now be determined starting with the given metric (\ref{21}) and applying the covariant transformation rules of the metric tensor, written in the form
\beq
g_{\mu\nu}(x)= {\pa x^{\prime \a}\over \pa x^\mu}
{\pa x^{\prime \b}\over \pa x^\nu} g'_{\a\b}(x').
\label{25}
\eeq
For additional checks we can also conveniently use the transformation of the Christoffel connection, given in general by: 
\beq
{\pa x^{\prime \a}\over \pa x^\mu}{\pa x^{\prime \b}\over \pa x^\nu}
\Ga'_{\a\b}\,^{\r} (x') = {\pa x^{\prime \r}\over \pa x^\la} \Ga_{\mu\nu}\,^\la (x)
- {\pa^2 x^{\prime \r}\over \pa x^\mu \pa x^\nu} .
\label{26}
\eeq
Expanding around the geodesic we can indeed exploit the properties of the FNC system, which imposes $\Ga'_{\a\b}\,^\r (x') =0$ at $x^{\prime A}=0$.

Let us first consider the component $(0,0)$ of Eq. (\ref{25}). By imposing that the metric transformation rule is satisfied up to the zeroth order and to the first order in $x$ we obtain, respectively, the conditions:
\beq
\dot \a^2 =1, ~~~~~~~~~~~~~~~~~~~~~~~~
\dot \b_k=0
\label{27}
\eeq
(a dot denotes a derivative with respect to $t$). Similarly, from the $(0,i)$ components evaluated up to the zeroth and to the first order in $x$, we get, respectively:
\beq
\b_k=0, ~~~~~~~~~~~~~~~~~~~ 
2 \ga_{ij} = \da_{AB}\, \dot \a^A\,_i\, \a^B\,_j.
\label{28}
\eeq

Let us finally consider the $(i,j)$ components of Eq. (\ref{25}), using the above result for $\b_k$. The zeroth order in $x$ gives the condition
\beq
A_{ij}=\da_{AB}\,  \a^A\,_i\, \a^B\,_j .
\label{29}
\eeq
This latter equation tells us that the $ \a^A\,_i$ can be thought of as ``triads" (the analog in three dimensions of the tetrads) for the zero-order metric $A_{ij}$. Assuming the latter to be non-degenerate, Eq. (\ref{29})  fixes the coefficient $\a^A\,_i$ modulo a (possibly time-dependent) rotation  of the form
\beq
\a^A\,_i \ra \widetilde{\a}\,^A\,_i= C^A\,_B \, \a^B\,_i,
\label{210}
\eeq
where the matrix $C^A\,_B $ satisfies the orthogonality condition
\beq
\da_{AB}\, C^A\,_C \, C^B\,_D = \da_{CD}
\label{211}
\eeq
(namely, $C^TC=I$). In addition, the $(i,j)$ components of Eq. (\ref{25}), to the first order in $x$, give the condition
\beq
B_{ijm}= \da_{AB}\left(  \a^A\,_i\, \b^B\,_{jm}+ \a^A\,_j\, \b^B\,_{im}\right).
\label{212}
\eeq

From Eq. (\ref{27}) we easily obtain $\a(t)=t$, modulo a sign and an unimportant additive constant. The time derivative of Eq. (\ref{29}), combined with (\ref{28}), gives $\dot A_{ij}= 4 \ga_{ij}$. We can thus rewrite the coordinate transformation (\ref{23}) in more explicit form as follows,
\beq
t'= t+ {1\over 4}  \dot A_{kl}(t) x^k x^l + \cdots, ~~~~~~~~~~~~~
x^{\prime A}= \a^A\,_k(t) x^k+ \b^A\,_{kl}(t) x^k x^l+\cdots ,
\label{213}
\eeq
where $\a^A\,_k$ satisfies Eq. (\ref{29}), and where $\b^A\,_{kl}$ is fixed in terms of $\a^A\,_k$ and of the first-order metric coefficient $B_{ijm}$, according to the following solution of Eq. (\ref{212}):
\beq
\beta^A\,_{jm} = \frac12 \alpha^{Ak} (B_{kjm} +B_{kmj}-B_{jmk}), ~~~~~~\alpha^{Ak} \equiv \a^A\,_m A^{mk}; ~~~~~~~A^{mk}A_{kn} = \delta^m_n. 
\label{sol212}
\eeq

The above results are  enough for the purpose of this paper, aiming at comparing the angular directions, locally defined in the FNC system around the wordline of the static geodesic observer, with the corresponding local directions defined in a different system of coordinates (the synchronous one, for the particular case at hand). 
The problem can be formulated as follows: given the definition of radial coordinate and angles around the observer's geodesic in the FNC system, such that
\beq
x'^A = r'  (\sin \theta' \cos \phi', \sin \theta' \sin \phi', \cos \theta')\; ,
\label{FNCangles}
\eeq
can we define a suitable coordinate transformation within the SG such that its angular variables can be identified with those of Eq. (\ref{FNCangles})?

We claim that, in general,  this is not possible. The crucial point here is that, because of  Eq. (\ref{29}), the relation between FNC and SG coordinates  depends explicitly on time if $A_{ij}$ does,  and therefore the would be coordinate transformation (which, in order to preserve the SG, should be a spatial diffeomorphism, see e.g. \cite{4a}) cannot define SG angles that coincide with those of the FNC system at all times. 
Indeed, this spatial diffeomorphism should have the form:
\beq
x^i  =x^i(r, \theta, \phi)\; ,
\label{SGdiff}
\eeq
and the only possibility to identify $\theta$ and $\phi$ with $\theta'$ and $\phi'$  at all times seems to be  the one offered by a FLRW homogeneous and isotropic metric, since in that case we can attribute all the time dependence in (\ref{213}) to the relation between $r'$ and $r$, while keeping the angles identified at all times. In order to check this, we can simply use the standard definition of polar coordinates:
\beq
x^i  = r (\sin \theta \cos \phi, \sin \theta \sin \phi, \cos \theta)\; ,
\label{SGdiffFRW}
\eeq
and bring the FLRW  metric to its well known SG form:
\beq
ds^2  = - dt^2 +a(t)^2 \left(\frac{dr^2}{1- kr^2}+ r^2 d \Omega^2\right) \; .
\label{SGdiffFRW1}
\eeq
Note that the spatial curvature parameter ($k = 0, \pm 1$) can be safely neglected around the $r=0$ geodesic.
One can then check that the above SG metric goes into the FNC metric (up to $O(r'^2)$ corrections)  with the identifications
\beq
r' = a(t) r~~;~~~~~~~~ \theta' = \theta ~~;~~~~~~~~ \phi' = \phi ~~;~~~~~~~~ t' = t +\frac12 a \dot{a}\, r^2\;, 
\label{ident}
\eeq
where the last relation between $t$ and $t'$ is nothing but the transformation of the time coordinates given by Eqs. (\ref{213}).

Note that we are {\it not} invoking a transformation involving $r$ as in (\ref{ident}), which would take us out of the SG: the needed residual gauge transformation is just (\ref{SGdiffFRW}).

It is relatively easy to see that such a procedure already fails to work in the (slightly more general) case of an anisotropic Bianchi I-type cosmology, 
 described in the synchronous gauge by the metric $g_{ij}=A_{ij}= a^2_i(t)\da_{ij}$ (no sum over $i$).
In such a case the general solution of Eq. (\ref{29}) gives $\a^A\,_j(t)=C^A\,_j \,a_j(t)$ (no sum over $j$), where $C^A\,_j$ are time-dependent coefficients satisfying Eq. (\ref{211}), and the coordinate transformation (\ref{213}) becomes
\beq
x'^A = \sum_k C^A\,_k \,(a_k x^k) + \cdots 
\label{216}
\eeq
 It is obvious that, unlike in the previous case, it is not possible now to introduce SG angular variables such that the r.h.s. of (\ref{216}) takes the standard form (\ref{SGdiffFRW}) up to a suitable definition of (a possibly time-dependent) $r$. The same considerations also apply to the so-called longitudinal gauge of cosmological perturbation theory\footnote{It is easy to show, for instance, that the longitudinal-gauge angles of a perturbed FLRW metric with tensor and scalar perturbations cannot be made to coincide at all times with the FNC angles, without going out of that gauge.}.

Of course, {\it at a given  time}, we can always perform a spatial reparametrization of the synchronous coordinates, $x^i \ra \tilde x^i(x)$
 which can make the angles coincide, but only at that chosen moment.
Let us finally notice  that the two sets of angles can also be made to coincide -- but, again,  only at a given time --  when a homogeneous and isotropic geometry is generalized by including scalar and tensor metric perturbations, to linear order, in the synchronous gauge. In that case,  we can always make the linear inhomogeneities  vanish at a given time, by exploiting the residual gauge freedom of the synchronous coordinates \cite{5}. This is also in agreement with the results recently presented in \cite{6}.

%%%%%%%%%%%%%%%%%%%%%%%%%%%%
%%%%%%%%%%%%%%%%%%%%%%%%%%
\section{Angular directions in the Fermi and GLC gauge}
\label{Sec3}
\setcounter{equation}{0}
%%%%%%%%%%%%%%%%%%%%%%%%%%

Let us now consider the class of cosmological geometries that can be described in terms of the GLC coordinates $y^\mu = (\tau, w, \theta^a)$, with $a=1,2$, by a metric of the form
\bea
\label{31}
&&
g_{\tau \tau}^{\rm{GLC}}=0, ~~~~~~~~~~~~~~~~~~~~~~~ g_{\tau w}^{\rm{GLC}} = - \up, ~~~~~~~~~~~~~ g_{\tau a}^{\rm{GLC}} =0, ~~~~
\\ \nonumber
&& 
g_{w w}^{\rm{GLC}} = \up^2 + \ga_{ab} U^a U^b, ~~~~~~ g_{w a}^{\rm{GLC}} =-U_a, ~~~~~~~~~~~~ g_{ ab}^{\rm{GLC}}= \ga_{ab},
\eea
where $U^a= \ga^{ab}U_b$, and $\ga^{ac} \ga_{cb} = \da^a_b$ (see e.g \cite{4}). 
It is important to note that a geodesic observer of this metric, with four-velocity $u_\mu= - \da_\mu^\tau$, exactly corresponds to a static geodesic observer of the synchronous gauge \cite{7} (specified, in our case, by the equation $x'^A= x^i=0$.)

For the purpose of this paper it will be important to work in the so-called ``temporal gauge" of the GLC coordinates \cite{8}, where the  past light-cones of a geodesic observer (i.e. the null hypersurfaces $w=$ const) are labelled by the reception time $\tau= \tau_o$ of the associated light signals.  It follows that, in such a gauge, we can set $\tau= w$ along the observer's geodesic, and we can expand the distance from it in power series of $(w-\tau)$. It can be shown \cite{8}, also, that the GLC  metric can always be normalized in such a way that  $- (g_{\tau w})_{w=\tau} = \up_{w=\tau} = 1$ along the geodesic worldline. 

We thus expand the transformation from the GLC to the Fermi coordinates, $y^\mu \ra x'^\alpha(y)$, around the static geodesic and in the temporal gauge, as follows :
\bea
&&
t'= \tau+ M_1(w, \theta) (w-\tau)^2 + \cdots , ~~~~~~~
\nonumber \\ &&
x'^A= N_0^A(w, \theta)  (w-\tau)+ {1\over 2} N_1^A(w,\theta)  (w-\tau)^2  + \cdots ,
\label{32}
\eea
where the positive quantity $(w-\tau)$ grows as one moves towards the past on a given light cone. The absence of the first-order term in the $t'$ transformation is consistent with the fact that the time $\tau$ coincides with the time coordinate $t$ of the synchronous gauge \cite{7}, and that -- as shown in Sect. \ref{Sec2} -- the Fermi time $t'$ differs from $t$ only at quadratic order in the distance from the geodesics. It also follows directly from imposing, on the geodesic, the GLC condition $g^{GLC}_{ww} = \up^2 + U^2$.

By also expanding in the limit $\tau \ra w$  the components of the  GLC metric \eqref{31} we now apply the covariant transformation rules of the metric tensor, which reads
\beq
g^{GLC}_{\mu\nu}(y)= {\pa x'^\a \over \pa y^\mu} {\pa x'^\b \over \pa y^\nu}g'_{\a\b}(x'),
\label{33}
\eeq
where $g'_{\a\b}(x')$ is the FNC metric (\ref{22}). By imposing 
$g_{\tau \tau}^{\rm{GLC}}=0$, according to Eq. \eqref{31}, we then obtain the condition
\beq
\da_{AB} N_0^A N_0^B=1
\label{34}
\eeq
to the zeroth order in $(w-\tau)$, and  the condition
\beq
\da_{AB} N_0^A N_1^B=-2M_1
\label{35}
\eeq
to the first order in $(w-\tau)$. By considering the transformation of $g_{\tau w}^{\rm{GLC}} = - \up$, and expanding $\up$ as
\beq
\up= 1+ \up_1(w,\theta) (w-\tau) + \cdots , ~~~~~~~~~~~~
\up_1 \equiv -\left(\pa \up\over \pa \tau\right)_{\tau=w}, 
\label{36}
\eeq
we find that the transformation rule is identically satisfied along the geodesics, while, to  first order in $(w-\tau)$, it gives the condition
\beq
2M_1= - \up_1.
\label{37}
\eeq

It is convenient, at this point, to choose our coordinate transformation (\ref{32}) in such a way that the two vectors $N_0^A$ and $N_1^A$ are proportional\footnote{Even if not explicitly assumed, the proportionality of $N_0^A$ and $N_1^A$ can be derived by combining the previous results with the transformations of the controvariant components of the GLC metric tensor.}, i.e. $N_1^A=K(w,\theta) N_0^A$. By inserting this ansatz into Eq. (\ref{35}), and using Eqs. (\ref{34}), (\ref{37}), we can immediately fix the proportionality coefficient to obtain
\beq
N_1^A= \up_1 N_0^A .
\label{38}
\eeq
As a consequence of this result (and of Eqs. (\ref{34}), (\ref{37})), it can be easily checked that the transformation rule (\ref{33}) automatically satisfies the condition $g_{\tau a}^{\rm{GLC}} =0$ not only along the geodesics, but also to first and  second order in the $(w-\tau)$ expansion.

Let us now apply the transformation (\ref{33}) to   $g_{w a}^{\rm{GLC}}$ and $g_{ab}^{\rm{GLC}}$. By expanding those metric components along the geodesic, and using the previous results for $N_0^A, N_1^A, M_1$, one finds that both $U_a$ and $\ga_{ab}$ are vanishing to the zeroth and  first order in the $(w-\tau)$ expansion. To second order they are non-vanishing, and can be expressed in term of the parameters $N_0^A$ as follows:
\bea
&&
U_a= -\left({1\over 2} \pa_a \up_1 + \da_{AB} \,\pa_w N_0^A \pa_a N_0^B \right) (w-\tau)^2 + \cdots ,
\label{39} \\ &&
\ga_{ab}= \da_{AB} \,\pa_a N_0^A \pa_b N_0^B \,(w-\tau)^2 + \cdots ,
\label{310}
\eea
where $\pa_w = \pa/\pa w$ and $\pa_a=\pa/\pa \theta^a$. Finally, for $g_{w w}^{\rm{GLC}}$, using Eqs. (\ref{34}) and (\ref{38}) we simply find:
\beq
g_{w w}^{\rm{GLC}} = \da_{AB} N_0^A N_0^B+2 \da_{AB} N_1^A N_0^B (w-\tau)+ \cdots = 1+2\Upsilon_1(w-\tau) +\cdots
\eeq
which agrees, to this order, with the GLC condition $g_{w w}^{\rm{GLC}}  =  \up^2 + \ga_{ab} U^a U^b$.

 Let us now show that, with an appropriate coordinate transformation allowed by the residual freedom of the GLC gauge \cite{5,8,9}, we can always choose the parameters $N_0^A$ in such a way that the angular coordinates of the GLC frame  coincide, at all times, with those determined by the angular directions measured by the Fermi geodesic observer as defined in Eq. (\ref{FNCangles}). 
 
In order to prove this assertion consider a (residual) gauge-fixing transformation $y^\mu \ra \wt y^\mu(y)$ for the GLC coordinates of the particular form
\beq
\wt \tau=\tau, ~~~~~~~~~~~ \wt w = w ,  ~~~~~~~~~~~ 
\wt \theta^a =\wt \theta^a (w, \theta),
\label{313}
\eeq
which still keeps us in the temporal gauge.
Let us choose, in particular, the following transformation:
\bea
&&
\theta^1 \ra \wt \theta^1 (w, \theta)= \arccos \left\{N_0^3\right\}, ~~~~~
\nonumber \\ &&
\theta^2 \ra \wt \theta^2 (w, \theta)= \arcsin \left\{N_0^2\left[
1- (N_0^3)^2\right]^{-1/2}\right\},
\label{314}
\eea
where, we recall,  $(N_0^1)^2+(N_0^2)^2+(N_0^3)^2=1$ according to Eq. (\ref{34}). By defining $\wt \theta^1= \wt \theta$ and $\wt \theta^2= \wt \phi$ one immediately obtains that, after such a gauge fixing, the relation (\ref{32}) takes the explicit leading order form
\beq
x'^A = N_0^A(\wt\theta)  (w-\tau)+ \cdots  
 \equiv (\sin \wt\theta \cos \wt\phi, \sin \wt\theta \sin \wt\phi, \cos \wt\theta)(w - \tau)+ \cdots ,
\label{315}
\eeq
which, compared with (\ref{FNCangles}),  brings to the identifications $r' = (w - \tau)$, $\theta' =  \wt\theta$,  $\phi' = \wt\phi$.

We can also check, for consistency, the behavior of the GLC metric around the geodesic in terms of the new coordinates $\wt y^\mu$ defined by Eqs. (\ref{313}) and (\ref{314})(see \cite{9} for the derivation of the whole set of gauge transformations). We then find that the component $g_{\tau w}^{\rm{GLC}} = - \up$ keeps unchanged under the given gauge transformation, so that $\wt \up = \up =1+ \up_1(w, \wt\theta)(w-\tau)+\cdots$. The metric components $\ga_{ab}$, on the contrary, are rescaled under the gauge fixing as follows
\beq
\ga_{ab} \ra \wt\ga_{ab}(w,\wt \theta)= \ga_{cd}(w,\theta)\,
{\pa \theta^c\over \pa \wt\theta^a}{\pa \theta^d\over \pa \wt\theta^b}.
\label{316}
\eeq
Hence, using Eqs. (\ref{310}) and (\ref{314}) we obtain (to leading order) the canonical form of the line-element on the two-sphere,
\beq
\wt\ga_{ab}(\wt \theta)= {\rm diag} \left(1, \sin^2 \wt\theta \right) (w-\tau)^2 + \cdots ,
\label{317}
\eeq
consistently with the choice of the FNC frame as the appropriate ``observational gauge" \cite{5,8,9}. Finally, by applying our gauge-fixing transformation to the metric component $g_{wa}^{\rm{GLC}}$, we have
\beq
U_a \ra \wt U_a(w,\wt\theta) = U_b\, {\pa \theta^b\over \pa \wt\theta^a}- \ga_{bc}\,
{\pa \theta^b\over \pa \wt w}{\pa \theta^c\over \pa \wt\theta^a}.
\label{318}
\eeq
By using the expansions (\ref{39}), (\ref{310}) for $U$ and $\ga$, and recalling that $\wt w= w$, $\pa N_0^i/\pa \wt w =0$, we obtain, to leading order:
\beq
\wt U_a(w,\wt\theta) = -{1\over 2} {\pa \up_1 \over \pa \wt \theta^a}\, (w-\tau)^2 + \cdots .
\label{319}
\eeq
This completes the form of the GLC metric, expanded around the 
geodesic, and gauge-fixed in such a way that its angular variables  $\wt \theta^a$ coincide with those of the FNC system. In this sense, ours can be seen as a concrete realization of the  \textit{observational gauge}\footnote{Note that  the behavior of $\tilde U_a$ and $\tilde \gamma_{ab}$ in (\ref{317}), (\ref{319}) implies that the controvariant vector $\tilde U^a$ does not vanish along the geodesic. This is in agreement with the fact that setting $U^a = 0$ on the geodesic corresponds to a different residual gauge fixing, the \textit{photocomoving gauge} \cite{8}, in which the angular coordinates do not coincide, in general, with the FNC ones.} whose existence was discussed in \cite{8}.

In order to illustrate the above results with a simple explicit example we may consider again the case of a Bianchi I-type geometry, described in the synchronous gauge by the metric $g_{ij}= a_i^2(t) \da_{ij}$. 

Let us first recall that the explicit transformation from the GLC to the synchronous    coordinates, $y^\mu \ra x^\mu(y)$, for such a geometry has already been derived \cite{8}, in the exact form and in the observational gauge, as follows:
\bea
&&
t= \tau, \nonumber \\ &&
x^i = x^i(\tau, w, \wt\theta)= a_i(w) N^i(\wt\theta) \int_\tau^w {d\tau' \over a_i^2(\tau')}
\left[ \sum_k {a_k^2(w)\over a_k^2(\tau')} N_k^2(\wt\theta)\right]^{-1/2}
\label{320}
\eea
(no sum over $i$). The unit vector $N^i(\wt\theta)$, which specifies the propagation direction of a light signal received  at the origin by a static geodesic observer,  takes exactly the same form as $N_0^A$  as a function of $\wt\theta$ in Eq. (\ref{315}) (and obviously satisfies $\da_{ij}N^iN^j=1$). By computing the corresponding metric components in the GLC frame \cite{8}, and expanding the result around the geodesic, we can then recover the leading terms in the typical form (\ref{36}), (\ref{317}), (\ref{319}) of the observational gauge, with
\beq
\up_1(w, \wt\theta)= -\sum_k N_k^2(\wt\theta) H_k(w),
\label{321}
\eeq
where $H_k = \dot a_k/a_k$.

Let us now recall that the transformation from the synchronous to the FNC system is described by Eq. (\ref{213}) where, for a Bianchi I-type geometry, we have $\dot A_{kl}= 2 a^2_k H_k \da_{kl}$, $\b^A\,_{kl}=0$ and $\a^A\,_k =\da^A_k a_k$ (no sum over $k$). Note that the rotational degrees of freedom, possibly associated with the matrix $C^A\,_B$ of Eq. (\ref{210}), have already been fixed, in our case, by the choice of the observational gauge when expressing the Bianchi metric in GLC coordinates (according to Eq. (\ref{320})). 

By expanding the transformation (\ref{320}) and the scale factor $a_k(t)$ around the geodesic, and inserting the results into Eq. (\ref{213}), we can then reconstruct the transformation connecting the GLC and FNC metric representations of our Bianchi geometry. To leading order we find
\bea
&&
t'= \tau+ {1\over 2} \sum_k N_k^2(\wt\theta) H_k(w) \,(w-\tau)^2 + \cdots , ~~~~~~~
\nonumber \\ &&
x'^A= N^A(\wt\theta) \, (w-\tau)+ {1\over 2} N^A(\wt\theta) \up_1(w,\wt\theta)
\, (w-\tau)^2  + \cdots ,
\label{322}
\eea
which exactly reproduces the transformation (\ref{32}) with the following gauge-fixed coefficients: $N_0^A = N^A(\wt\theta)$, $N_1^A= N^A \up_1$, $M_1= -\up_1/2$, and  with $\up_1$ specified by Eq. (\ref{321}). 
The (leading order) relation between the angular directions thus reduces to $n'^A= N^A$, and the time dependence of the Bianchi metric fully disappears from this relation, 
Hence, unlike the angular directions defined by the synchronous system of coordinates, those of the GLC system can be safely identified with the angles measured by a static geodesic observer, and not only for an isotropic geometry (as in the case of the synchronous gauge).

This implies, in particular, that the GLC coordinates can be conveniently used to compute observable dynamical effects like -- for instance -- the redshift drift effect (see e.g. the detailed analysis presented in \cite{4}), while the same computations performed in the synchronous gauge cannot be directly compared with the observational data.

Our consistency checks can be extended to higher orders in the distance from the geodesic. This exercise is carried out in detail, up to quadratic level, in the Appendix.

%%%%%%%%%%%%%%%%%%%%%%%%%%%%
%%%%%%%%%%%%%%%%%%%%%%%%%%%%%
\section{Summary and a claim}
\label{Sec4}
\setcounter{equation}{0}
%%%%%%%%%%%%%%%%%%%%%%%%%%%%

A simple way to summarize the results of this paper is that it completes the constructive approach \cite{8} to defining GLC coordinates by essentially completely gauge fixing them.

The residual gauge freedom $w \ra \tilde{w}(w)$ can be fixed by requiring $w$ along a given past light cone to coincide with the proper time $\tau$ of the observer at the tip of the cone. This is the temporal gauge discussed in \cite{8}. Here we have shown that, thanks to the other residual gauge freedom \cite{5,9}, $\theta^a \ra \wt{\theta}^a(w,\theta)$, it is also possible to define the GLC angles along the observer's geodesic to be those of the Fermi coordinates along that same geodesic. The global definition of the GLC angles then simply consists in saying that they are constant (modulo the occurrence of caustics? See e.g. \cite{10}) along the null rays  lying on each light cone.

We would like now to the address the question (see \cite{9}) of whether one can already identify the observational (Fermi) angles by just  working inside the GLC system, i.e. without constructing, case by case, the corresponding Fermi coordinates. In this connection we may note that Eq. (\ref{319}) (omitting the tilde for simplicity of notation), rewritten as
\beq
\pa_{\tau} U_a = \pa_a \up + O((w-\tau)^2), 
\eeq
namely
\beq
U_a = -\int_{\tau}^w \pa_a \up(\tau',w, \theta^a) d \tau'  + O((w-\tau)^2),
\label{UY}
\eeq
should provide, within the GLC,  the necessary and sufficient condition for the above-mentioned identification of the angular variables\footnote{Note that Eq. \eqref{UY} already implies that $\pa_w U_a=-\pa_a\up+O((w-\tau)^2)$. It therefore corresponds to just two conditions, to be fulfilled via a residual gauge transformation in the GLC coordinates.}.

In order to assess the validity of this claim we observe that, assuming Eq. (\ref{UY}),  a straightforward calculation allows us to rewrite the GLC metric, 
around the geodesic and in terms of a new time parameter $ \tilde{\tau}$,
in the particularly simple form:
\bea
&&
ds^2 = -2 dw d\tilde{\tau} + dw^2 + \ga_{ab} d\theta^a d\theta^b +  O((w-\tau)^2) dw^2 ,
\nonumber \\ &&
 \tilde{\tau} \equiv  w  -\int_{\tau}^w  \up(\tau',w, \theta^a) d \tau'  \, ,
\label{ds2}
\eea
where we recognize in $\tilde{\tau}$ just the expression of the Fermi time $t'$ as given by Eqs. (\ref{32}), (\ref{36}) and (\ref{37}), up to higher order curvature corrections.

Recalling now that, in an infinitesimal neighborhood of the geodesic (in practice at distances much smaller than the curvature radius of the geometry), the metric is flat, we conclude that $\ga_{ab}$, as a function of $\tilde{\tau}, w, \theta^a$, must be equivalent (up to diffeomorphisms) to the metric of a sphere of radius $r' = (w-\tilde{\tau}) \sim (w- \tau)$, i.e.
\beq
\ga_{ab} = \hat{\ga}_{ab}(w, \theta^a)(w- \tau)^2 + O((w-\tau)^3) \; ,
\label{gahat} 
\eeq 
with $\hat{\ga}_{ab}$ the metric of the unit sphere, possibly written in some complicated $w$-dependent angular coordinates. But then, transforming it back to the standard form (\ref{317}), would induce a shift in $U_a$, thus contradicting (\ref{UY}). Since we have shown that (\ref{UY}) and (\ref{317}) can and should be simultaneously satisfied, we  conclude that $\ga_{ab}$ must already take the form  (\ref{317}) modulo a global (i.e. $w$-independent) angular reparametrization. We thus claim that Eq. (\ref{UY}) does indeed characterize the observational gauge within the GLC  coordinate system.

%%%%%%%%%%%%%%%%%%%%%%%%%%%%%%%%%%%%%%%%%%%%%%%

%%%%%%%%%%%%%%%%%%%%%%%%%
\vskip 1 cm
 
\section*{Acknowledgements}
We are grateful to Pierre Fleury and Ermis Mitsou for useful comments on a preliminary draft of this paper. GF thanks Fulvio Scaccabarozzi for helpful discussions during the early stages of this project and GV would like to acknowledge an instructive discussion with Misao Sasaki. GF, MG and GM are supported in part by INFN under the program TAsP (Theoretical Astroparticle Physics). GF is supported by the Swiss National Science Foundation and by a Consolidator Grant of the European Research Council (ERC-2015-CoG grant 680886). We wish to thank the hospitality and financial support of the Dipartimento di Fisica and Sezione INFN di Pisa, where an important part of this work has been carried out.

\appendix
\section{Second-order metric transformation and further consistency checks}

In this Appendix we shall apply the transformation (\ref{33}) up to the second order in the distance from the geodesic, on all components of the GLC metric, extending in this way the leading-order results already obtained in Sect. \ref{Sec3}. As expected from Eqs. (\ref{22}), (\ref{Riem}), it will be shown that the Riemann corrections must be included into the second-order computations, and play a crucial role for the consistency of the transformation between the GLC and FNC system and the related identification of their angular variables. 

Let us start with the appropriate higher-order generalization of the coordinate transformation (\ref{32}), which we write in the form
\bea
&&
t'=\tau + M_1(w-\tau)^2+M_2(w,\theta)\, (w-\tau)^3,
\nonumber\\ &&
x'^A=N_0^A\,(w-\tau)+\frac{1}{2}\,N_1^A\,(w-\tau)^2+\frac{1}{6}\,N_2^A(w,\theta) \,(w-\tau)^3,
\label{a1}
\eea
where the functions $M_1$ and $N_1^A$ are those already determined in Sect. \ref{Sec3}. Also, let us first consider the case of the observational gauge, with $N_0^A= N_0^A(\theta)=\left( \sin\theta\cos\phi,\sin\theta\sin\phi,\cos\theta \right)$
(for the sake of simplicity, we shall omit the tilde on the angular variables). Finally, let us denote with $f_{\a\b}$ the second-order contributions to the transformation (\ref{33}) arising from the curvature corrections intrinsic to the Fermi metric, and define $f_{\a\b} \equiv f_{\a A \b B} N_0^A N_0^B$. One then obtains, from Eq. (\ref{Riem}), 
\beq
f_{00}=\left(R'_{0A0B}N_0^AN_0^B\right)_{w=\tau}, ~~~
f_{0C}=\frac{2}{3}\left(R'_{0ACB}N_0^AN_0^B\right)_{w=\tau}, ~~~
f_{CD}=\frac{1}{3}\left(R'_{CADB}N_0^AN_0^B\right)_{w=\tau}.
\label{a2}
\eeq

By applying Eq. (\ref{a1}) to the metric transformation (\ref{33}), and using the symmetry properties of the Riemann tensor (in particular, $R_{\mu\nu\a\b}= R_{[\mu\nu][\a\b]}$), we then find that the condition $g_{\tau \tau}^{\rm{GLC}}=0$ implies that the higher-order coefficients $M_2$, $N_2^A$ must be related by
\beq
6M_2= -\left( f_{00}+ \da_{AB} N_0^A N_2^B \right).
\label{a3}
\eeq
Similarly, by expanding to second order $Y$ as
\beq
\Upsilon=1+\Upsilon_1(w, \theta)(w-\tau)+\Upsilon_2(w, \theta)(w-\tau)^2+\cdots
\label{a4}
\eeq
the computation of $g_{\tau w}^{\rm{GLC}}$ leads to the condition
\beq
\up_2 = 3 M_2 +\da_{AB} N_0^A N_2^B.
\label{a5}
\eeq
Let us then compute the component $g_{w w}^{\rm{GLC}}$, up to second order, in the observational gauge. We obtain
\beq
g_{w w}^{\rm{GLC}}= 1 +2 \up_1 (w-\tau) + \left(\da_{AB} N_0^A N_2^B+ \pa_w \up_1 \right) (w-\tau)^2 + \cdots.
\label{a6}
\eeq
A generic GLC metric, on the other hand, must satisfy the condition $g_{w w}^{\rm{GLC}}=\up^2+ \ga^{ab} U_a U_b$ (see Eq. (\ref{31})). By using the form (\ref{317}), (\ref{319}) of the metric components in the observational gauge we can expand $U^2$ up to second order as
\beq
U^2 \equiv \ga^{ab} U_a U_b = {1\over 4} \ga_0^{ab} \pa_a \up_1 \pa_b \up_1 (w-\tau)^2 + \cdots,
\label{a7}
\eeq
where $ \ga_0^{ab}$ is simply the inverse of the diagonal metric coefficient $\ga^0_{ab}= \rm{diag} (1, \sin^2 \theta)$ appearing in Eq. (\ref{317}). Hence, by using for $\up$ the expansion (\ref{a4}), and comparing with the result (\ref{a6}), we find that the condition $g_{w w}^{\rm{GLC}}=\up^2+ U^2$ is satisfied up to second order provided that
\beq
\up_1^2 +2 \up_2 +{1\over 4} \ga_0^{ab} \pa_a \up_1 \pa_b \up_1=
\da_{AB} N_0^A N_2^B+ \pa_w \up_1.
\label{a8}
\eeq
Using Eq. (\ref{a5}) for $\da_{AB} N_0^A N_2^B$ and eliminating $M_2$ through  Eq. (\ref{a3}) the above condition reduces to
\beq
f_{00}= \up_1^2 +{1\over 4} \ga_0^{ab} \pa_a \up_1 \pa_b \up_1 - \pa_w \up_1.
\label{a9}
\eeq

In order to check the internal consistency of such condition let us now compute the curvature contributions $f_{00}$ for a  geometry described by a generic GLC metric (\ref{31}). Let us start with the definition (\ref{a2}) of $f_{00}$, and express the Riemann tensor $R'_{\mu\nu\a\b}(x')$ in terms of the corresponding GLC components, $R_{\mu\nu\a\b}(y)$, through the coordinate transformation (\ref{a1}). We obtain, on the geodesic,
\beq
f_{00}=\left(R'_{0A0B}N_0^AN_0^B\right)_{w=\tau}=
\left( R^{\a\mu\b\nu} \pa_\a t'\pa_\mu x'^A \pa_\b t'\pa_\nu x'^B\right)_{w=\tau}N_{0A} N_{0B},
\label{a10}
\eeq
so that, by applying the transformation (\ref{a1}) and the normalization (\ref{34}), 
\bea
f_{00}&=& \left(R^{\tau\mu\tau\nu}\right)_{w=\tau}\left( \da_\mu^\tau- \da_\mu^w\right)
\left( \da_\nu^\tau- \da_\nu^w\right) =  \left(R^{\tau w \tau w}\right)_{w=\tau} = \left(g^{\tau \a} g^{w \mu} g^{\tau \b} g^{w \nu} R_{\a\mu\b\nu} \right)_{w=\tau}
 \nonumber \\ &=&
 \left(R_{w\tau w\tau} + 2 R_{w\tau a\tau} U^a + R_{a\tau b\tau} U^a U^b\right)_{w=\tau}.
 \label{a11}
 \eea
An explicit computation for the metric (\ref{31}) then shows that   the last two terms -- on the geodesic and in the observational gauge -- are identically vanishing, while the first term gives, as the only nonzero contribution,
\beq
f_{00} =  \left(R_{w\tau w\tau}\right)_{w=\tau}=  \left(g_{\tau \a}R_{w\tau w}\,^\a\right)_{w=\tau}= \left(-\up R_{w\tau w}\,^w\right)_{w=\tau} 
=\left(\up \pa_\tau \Ga_{ww}\,^w\right)_{w=\tau},
\label{a12}
\eeq
where
\beq
 \Ga_{ww}\,^w = \pa_\tau \up +{\pa_w \up\over \up} + {1\over 2} {\pa_\tau U^2\over \up}
 \label{a12a}
 \eeq
(see e.g. \cite{12} for an explicit computation of the Christoffel connection for the GLC metric). We thus obtain the curvature contribution
\beq
f_{00} =\left(\up \pa_\tau \Ga_{ww}\,^w\right)_{w=\tau}
=\up_1^2 +{1\over 4} \ga_0^{ab} \pa_a \up_1 \pa_b \up_1 - \pa_w \up_1,
\label{a12b}
\eeq
and comparing the above result with eq. (\ref{a9}) we may  conclude that the condition $g_{w w}^{\rm{GLC}}=\up^2+ U^2$ is always satisfied up to second order, in the observational gauge, for any given GLC metric.

The same consistency check can be performed even outside the observational gauge, considering a generic parametrization with $N_0^A = N_0^A (w,\theta)$. In that case, in the computation of $g_{w w}^{\rm{GLC}}$ we obtain additional second-order contributions from the more general form of $U^a$ (see Eq. (\ref{310})), but there are also new metric contributions to the Christoffel connection and to the components of the Riemann tensor. It can be checked that, when both types of contributions are correctly included, they exactly cancel each other, and the condition $g_{w w}^{\rm{GLC}}=\up^2+ U^2$ keeps to be satisfied, to the second order around the geodesic, quite independently of the gauge fixing prescription.

Let us conclude this Appendix with the explicit example of the Bianchi I-type geometry, which can be expressed in GLC coordinates and in the observational gauge, as discussed in \cite{8}, with $N_0^A= \da^A_i N^i(\theta)$ and with $\up_1(w,\theta)$ given by Eq. (\ref{321}) (see also Sect. \ref{Sec3}).
 
To check the validity of Eq. (\ref{a9}) for such a geometry let us start with the curvature contribution $f_{00}$, and let us express the Fermi components of the Riemann tensor, $R'_{\mu\nu\a\b}(x')$,  in terms of its synchronous-gauge components, related to the Fermi ones by the coordinate transformation (\ref{213}). 
As discussed in Sects. \ref{Sec2}, \ref{Sec3}, the coefficients $\a^A_k$ appearing in Eq. (\ref{213}) for the Bianchi I geometry are given by $\a^A_k = \da^A_k a_k(t)$ (no sum over $k$). It follows that, on the geodesic, $\pa t'/\pa t=1$, $\pa x'^A/\pa x^i= \da^A_i a_i$, and we obtain
\beq
f_{00}=\left(R'_{0A0B}N_0^AN_0^B\right)_{x'^A=0}=
 \da^i_A a_i^{-1} \,
 \da^k_B a_k^{-1} N_0^AN_0^B R_{0i0k}
 =\sum_{i,k} \left(N^i\over a_i\right)\left(N^k\over a_k\right)R_{0i0k},
 \label{a13}
\eeq
where $R_{0i0k}$ are the components Riemann tensor for the Bianchi geometry computed in synchronous coordinates, i.e.
 \beq
R_{0i0k} = \da_{ik} a_k^2 \left(\dot H_k +H_k^2\right).
\label{a14}
\eeq
The curvature contribution to Eq. (\ref{a9}) takes thus the form
\beq
f_{00}= \sum_k N_k^2 \left[\dot H_k(w) +H_k^2(w)\right].
\label{a15}
\eeq

Let us now compute the right-hand side of Eq. (\ref{a9}). By recalling that $\up_1$
is given by Eq. (\ref{321}) we have
\beq
-\partial_w\Upsilon_1 = \sum_k N_k^2 \dot H_k,
\label{a16}
\eeq
and
\beq
\frac{1}{4}\gamma_0^{ab}\partial_a\Upsilon_1\partial_b\Upsilon_1
\equiv \gamma_0^{ab} \sum_{k,j}N_k \left(\pa_a N_k \right)N_j\left(\pa_b N_j\right) H_k H_j.
\label{a17}
\eeq
On the other hand, by using the definition  $\ga^0_{ab}= \pa_a N_k \pa_b N_j \da^{ij}$ (see Eq. (\ref{310})), we have
\beq
 \gamma_0^{ab} \pa_a N_k \pa_b N_j = \da_{kj} -N_kN_j,
 \label{a18}
 \eeq
so that Eq. (\ref{a17}) can be rewritten as
\beq
\frac{1}{4}\gamma_0^{ab}\partial_a\Upsilon_1\partial_b\Upsilon_1
=\sum_k N_k^2  H_k^2 -\left(\sum_k N_k^2  H_k\right)\left(\sum_j N_j^2  H_j\right) \equiv 
\sum_k N_k^2  H_k^2 - \up_1^2.
\label{a19}
\eeq
Inserting this result, together with Eq. (\ref{a16}), into the right-hand side of Eq. (\ref{a9}) we can exactly reproduce the curvature contribution (\ref{a15}), and thus satisfy the required condition for the consistency of the GLC gauge.

%%%%%%%%%%%%%%%%%%%%%%%%%%%%
%%%%%%%%%%%%%%%%%%%%%%%%%%%%%%

\end{document}